\pdfoutput=1
\documentclass[a4paper]{jpconf}
\usepackage{graphicx}
\usepackage{amsmath}
\usepackage[utf8]{inputenc}
\usepackage{graphicx}
\usepackage{subcaption}
\usepackage[colorlinks = true,
            linkcolor = blue,
            urlcolor  = blue,
            citecolor = blue,
            anchorcolor = blue]{hyperref}

\begin{document}
\title{Shock waves in Interstellar Cloud-Cloud and Wind-Cloud Collisions}

\author{S.~Navarrete$^{1}$, B.~J.~Pinargote$^{1}$, and W.~E.~Banda-Barrag\'{a}n$^{1,2}$}

\address{$^{1}$ Escuela de Ciencias F\'isicas y Nanotecnolog\'ia, Universidad Yachay Tech, Hacienda San Jos\'e S/N, 100119 Urcuqu\'i, Ecuador\\
$^{2}$ Hamburger Sternwarte, University of Hamburg, Gojenbergsweg 112, 21029 Hamburg, Germany\\}

\ead{wbandabarragan@gmail.com}

\begin{abstract}
The interstellar medium (ISM) is a key ingredient of galaxies  and their evolution, consisting of multiphase, turbulent dust and gas. Some of the star-forming regions in our Galaxy originate from cloud-cloud and wind-cloud collisions, which generate shock waves that change the physical and chemical properties of the gas. We utilise our own python-based shock-finding algorithm to study the properties and distribution of shocks in interstellar collisions. Such interactions are studied via 3D numerical simulations with different initial conditions: Cloud-cloud collisions (CCc): We identify four stages of evolution: pre-collision, compression, pass-through, and dissipation. We also vary the size of one of the colliding clouds. Larger clouds facilitate cloud erosion and the formation of more and stronger shocks at early stages. Shock distributions are also time-dependent, as strong shocks are only produced during the early stages. As the collisions evolve, turbulent kinetic energy is rapidly dissipated, so most perturbations become subsonic waves at late times. Wind-cloud collisions (WCc): we identify four stages: compression, stripping, expansion, and break-up. We study the evolution of several diagnostics in these clouds: energies (thermal and kinetic), temperature, displacement of the centre of mass, and mass-weighted averages of the cloud density and acceleration. We show, that the geometry of the cloud impact the diagnostic parameters, for example, smoothing the edges of the cloud leads to enhanced mass losses and dispersion, but has little impact on the shock distribution.
\end{abstract}

\section{Introduction} 

The interstellar medium (ISM) is a complex component of galaxies. It comprises gas, dust, comic rays, electromagnetic radiation, magnetic fields, gravitational fields, and dark matter \cite{2011piim.book.....D}. All these components fill the space between stars and interact with them, making the ISM a dynamic environment. The ISM’s dynamic nature is controlled by various processes, such as ionization \cite{2019A&A...622A.143C}, electromagnetic radiation\cite{2015A&A...582L..13P}, star formation (SF) \cite{2014MNRAS.444.1301P}, turbulence \cite{2023PASA...40...46K}, and cloud collisions \cite{2013ApJ...774L..31I}. Star formation is the most relevant process of ISM dynamics because it fuels the new star, which can be triggered by cloud-cloud collisions and wind-cloud collisions \cite{2013ApJ...774L..31I} \cite{2019A&A...622A.143C}. Besides such stellar feedback can launch atomic and molecular outflows e.g. \cite{2023A&A...674L..15V}.\par

Cloud-cloud and wind-cloud collisions are processes strongly linked with star formation and ISM evolution \cite{2013ApJ...774L..31I} \cite{2012A&A...541A..63P}. Both processes affect the physical and chemical properties, and the structure of the ISM. Cloud-cloud collisions occur when two gas clouds interact, perturbing some regions of the ISM. The effects of the collision can drive the formation of new stars \cite{2015MNRAS.453.2471B} \cite{2015MNRAS.450...10H}\cite{2017ApJ...835..142T}. Wind-cloud collision is more related to feedback processes. As stellar winds, produced by SF propagate, they encounter and interact with ISM clouds, leading to the formation of shells \cite{2020MNRAS.499.2173B} and filaments \cite{2018MNRAS.473.3454B}.\par

Observations of cloud-cloud collisions (CCc) have targeted density tracers (such as CO isotopologues) and shock tracers (such as $\rm{SiO}$, $\rm{HNCO}$, $\rm{CH_3OH}$). By studying their different transitions \cite{2020MNRAS.499.4918A}\cite{2020MNRAS.497.4896Z}, three main observational features characterize CCc: a) A complementary density distribution between gas at low and high velocities, b) a gas bridge in position-velocity space, and c) a U-shape \cite{2013ApJ...774L..31I}. The candidate ISM regions for hosting CCc are zones with high SF rates, mostly located in the Galactic center and stellar disk of our Galaxy \cite{2018ApJ...853..171G}\cite{2018ApJ...859..166F}.\par

Moreover, observational studies indicate that high-speed winds interact with atomic and molecular clouds. In areas of star formation, high-velocity winds and shocks can significantly impact molecular clouds. For example, the B59 filament in the Pipe Nebula is believed to be compressed and distorted by a wind \cite{2012A&A...541A..63P}. Other wind-cloud interactions can be seen in morphological features of ISM clouds. For instance, cloud disruption, fragmentation, and dispersion can be seen in the Rosette molecular cloud \cite{2010ApJ...719.1872B}. Another outcome of these interactions involves the incorporation of the cloud into the stellar-driven winds, resulting in the production of shock waves \cite{2021MNRAS.506.5658B} and the acceleration of the cloud material in ISM flow \cite{2012ApJ...761L..21S} \cite{2012ApJ...751...69S}.\par

In this paper, we delineate two types of simulations: cloud-cloud collisions (CCs) and wind-cloud collisions (WCs). Our objectives are as follows: Understand how changes in the initial conditions in three-dimensional, quasi-isothermal, hydrodynamical simulations of cloud-cloud collisions, such as the diameter of the clouds, affect the physical properties of the gas; Identify gas perturbations in both colliding clouds, and characterize them into shock waves (Mach number, ${M>\,1}$) and subsonic waves ($M<\,1$); Study the evolution of various diagnostics, namely: total mass, energy densities (thermal and kinetic), temperature, displacement of the centre of mass, mass-weighted velocity, acceleration and shock distribution of clouds in wind-cloud models.


The paper has the following parts: in Section \ref{ISMshocks} we present a summary of interstellar shock theory, in Section \ref{Sims} we show our simulation setups, in Section \ref{results} we show the simulations results, and in Section \ref{conclusions} we list our conclusions.

\section{Interstellar shock theory}
\label{ISMshocks}
First, we briefly summarise the theory of shocks. Using the Euler equations \cite{landau2013fluid}, we derive the simplified Rankine Hugoniot (RK) jump conditions\cite{constantin2005euler}, which describe the behavior of gas across shock discontinuities as a function of the specific heat capacity ratio $\gamma$ and the shock Mach number $M= \frac{v}{c_s}$ which describes how fast the perturbation is compared to the sound speed in the medium.\par

Shock waves are perturbations in gases, characterized by a sudden change in physical properties (i.e., by jumps in temperature $T$, density $\rho$, thermal pressure $p$). Shocks move faster than the sound speed in the medium, so by definition, they have a Mach number larger than 1 ($M > 1$). The RK jump conditions \cite{constantin2005euler} allow us to calculate the shock Mach numbers in simulation grid cells as $M_i = \sqrt{ {M_{x}}^2 + {M_{y}}^2 + {M_{z}}^2}$, where each component is obtained from the directional speed gradient $\nabla v_{x,y,z} \approx \left| \frac {\partial{v}}{\partial x,y,z}\right|(2 \Delta x,y,z)$, where the derivatives can be approximated by the central- difference method.
\begin{equation}
 M_{x,y,z} = \frac{-\nabla v_{x,y,z}(1+\gamma)+\sqrt{16 {c_s}^2+{\nabla v_{x,y,z}}^2(1+\gamma)^2}}{4 c_s}
 \label{Mach_eq}
\end{equation}
where $c_{s}$ is the sound speed  $c_{s} = \sqrt{\gamma \frac{P}{\rho}}$ for $P$ and $\rho$ are the thermal pressure and density. Gas perturbations with $M > 1$ are shock waves, the ones with $M=1$ are transonic waves, while subsonic waves have Mach numbers $M<1$.\par

In addition, collisions between two clouds or between a cloud and a fast-moving wind can induce turbulence, even if we do not explicitly include turbulence in the clouds or the wind \cite{2019MNRAS.486.4526B}. To quantify it, we use the $M_{\rm rms}$, which is the ratio between the velocity dispersion $\sigma_v$ and the sound speed $c_s$  then :
\begin{equation}
    M_{\rm rms} = \frac{\sigma_v}{c_s}
\end{equation}


\section{Simulations}
\label{Sims}
Numerical simulations are powerful tools for studying shock waves because they allow the exploration of a wide parameter space. There are different approaches to simulating events that produce shock waves. In this case, to perform numerical simulations, we use the PLUTO code \cite{2007ApJS..170..228M} to solve the Euler equations. All our models were performed using the hydrodynamics module in a 3D Cartesian coordinate system (X, Y, Z). The analysis of the simulation output files 
was carried out using Python 3.8 \cite{python}, as it is a flexible, open-source tool that suited the type of analysis we were interested in.

We simulate two different models (Habe-Ohta (H-O) model \cite{2014ApJ...792...63T} and identical cloud (I-C) model\cite{1970ApJ...159..277S}\cite{1970ApJ...159..293S}) of the collision of two clouds in the interstellar medium, where one moves to the other with a certain velocity. In the H-O model, where the moving cloud is small compared with the initially non-moving cloud, we vary the initial velocity of the moving cloud on a range from $\rm {20}$  to $\rm{90\, km \,s ^{-1}}$. In this paper we present models with a collision velocity of $\rm{30km s^{-1}}$. For additional CCc models at other velocities, see \cite{4480}. In the I-C model, the two clouds are the same size. We consider a quasi-isothermal $\gamma = 1.01$ and a mean particle mass $\rm{\mu =2.36}$  to mimic the conditions in the interstellar molecular gas \cite{{2011MNRAS.412..469A}} \cite{1986ApJ...305..309H}. The initial conditions were constrained by Larson's laws. Table \ref{tab:cc_models} shows the initial condition of both models. 

    
\begin{table}[!htbp]
    \centering
    \begin{tabular}{c c c c c c}
        \br
        \textbf{Simulation ID} & \multicolumn{2}{c}{\textbf{Cloud Diameter}} & \multicolumn{2}{c}{\textbf{Cloud Initial Velocity}} & \textbf{Polytropic Index}\\
         & $\, C_1\, \rm{(pc)} \, $  & $\, C_2\,\rm{(pc)}\,$ &$\, C_1\,\rm{km \,s^{-1})} \, $  & $\, C_2\, \rm{(km \,s^{-1})} \,$ & $\rm{\gamma}$ \\
         \hline
       H-O model & 2 & 4  & $30 $ & $0 $ &1.01 \\
        I-C model & 4 & 4 & $30 $ & $0 $ &1.01 \\ 
        \br
    \end{tabular}
   \caption{Initial conditions of cloud-cloud collisions. Cloud 1 represents the moving cloud, while Cloud 2 shows the initially non-moving cloud values.}
    \label{tab:cc_models}
\end{table}

\begin{figure}[h!]
    \centering
    \begin{subfigure}[b]{0.45\linewidth}
        \includegraphics[width=\linewidth]{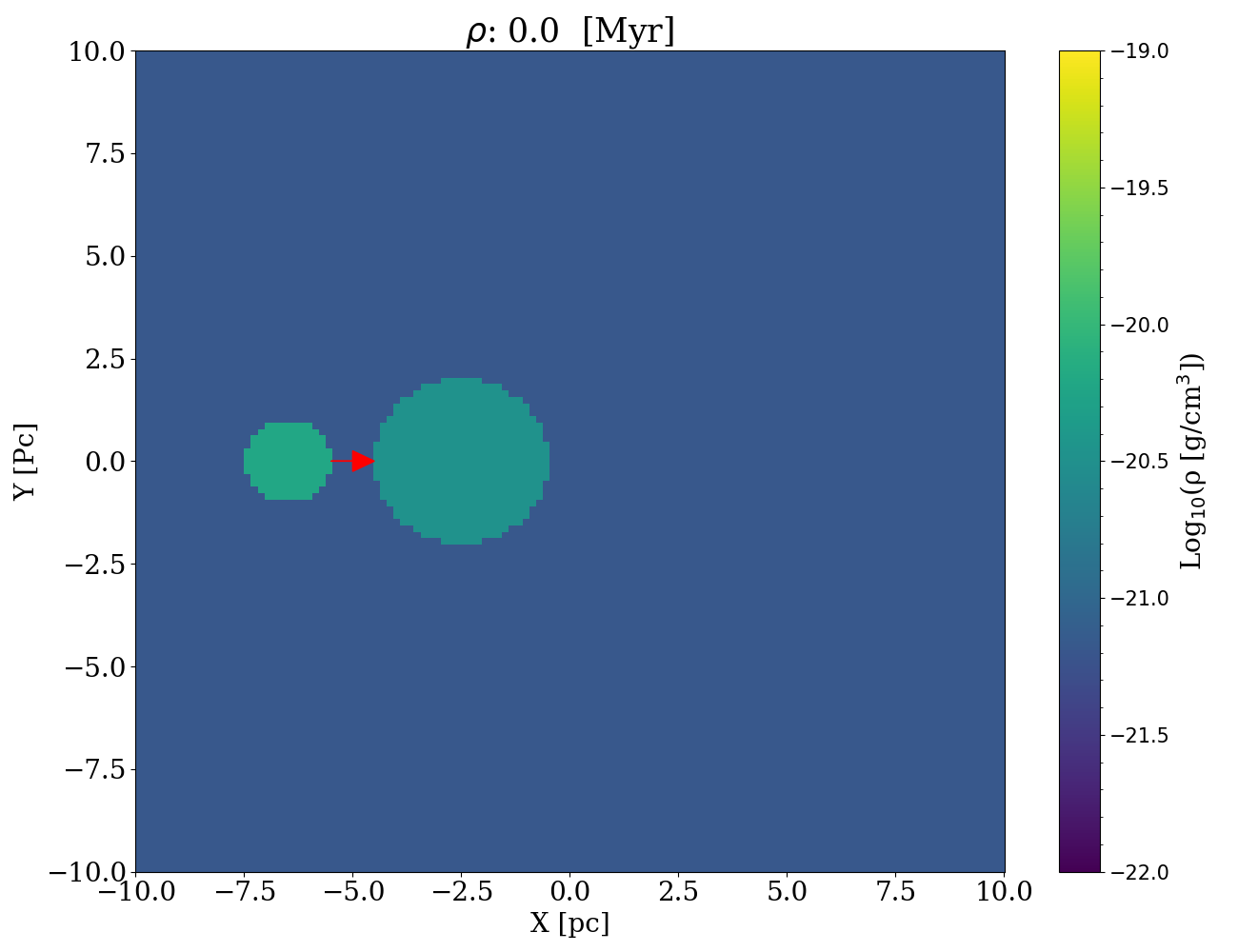}
        \caption{Habe-Ohta model.}
        \label{fig:HB model}
        \end{subfigure}
        \begin{subfigure}[b]{0.45\linewidth}
        \includegraphics[width=\linewidth]{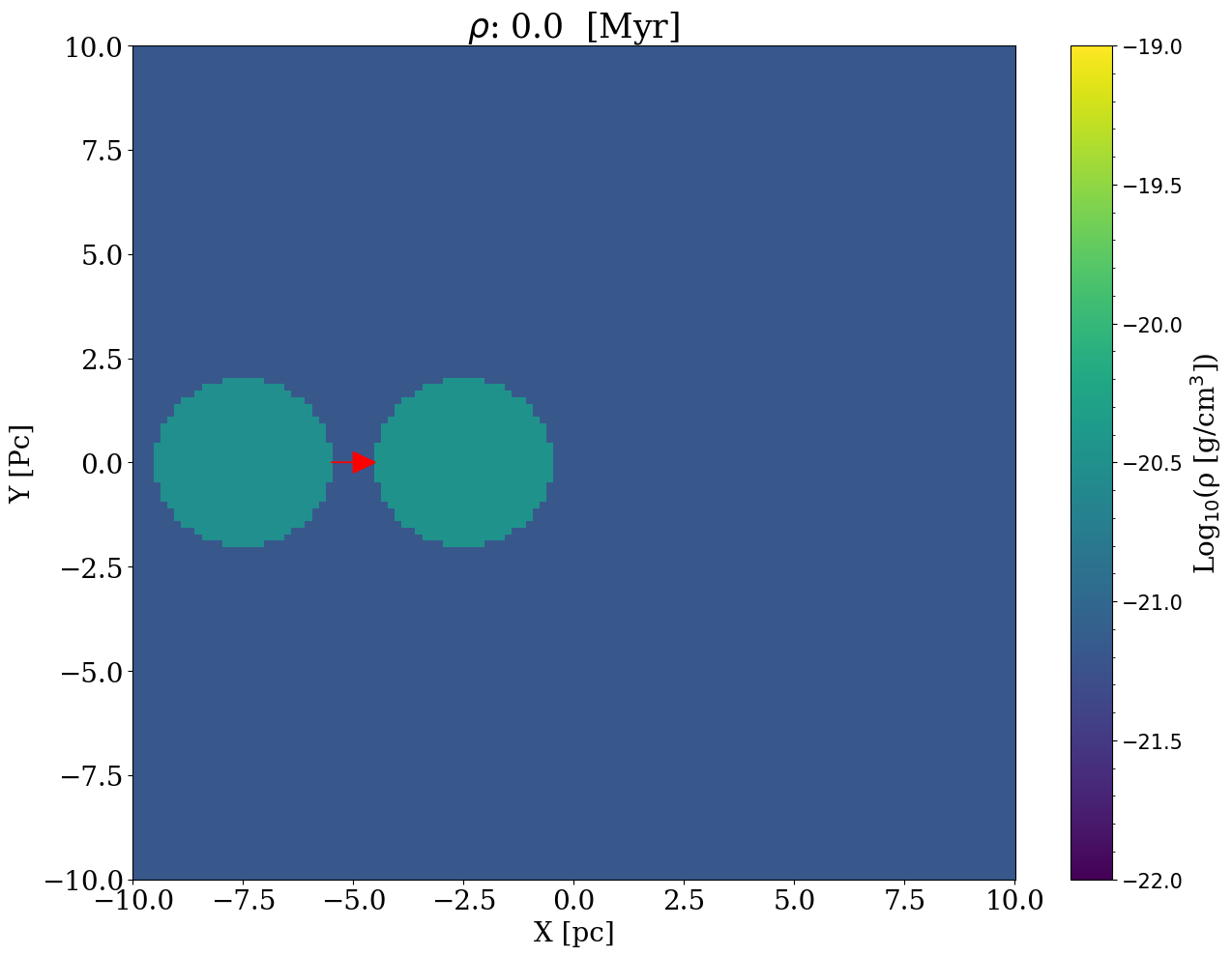}
        \caption{Identical cloud model.}
        \label{fig:Ic model}
    \end{subfigure}
    \caption{2D Density slices on the XY plane at $z=0$ and $\rm t = 0.0$ Myr of the two cloud-cloud models. The red arrow show the collision direction.\it}
    \label{fig:CCc}
\end{figure}

\begin{figure}[h!]
    \centering
    \begin{subfigure}[b]{0.3\linewidth}
        \includegraphics[width=\linewidth]{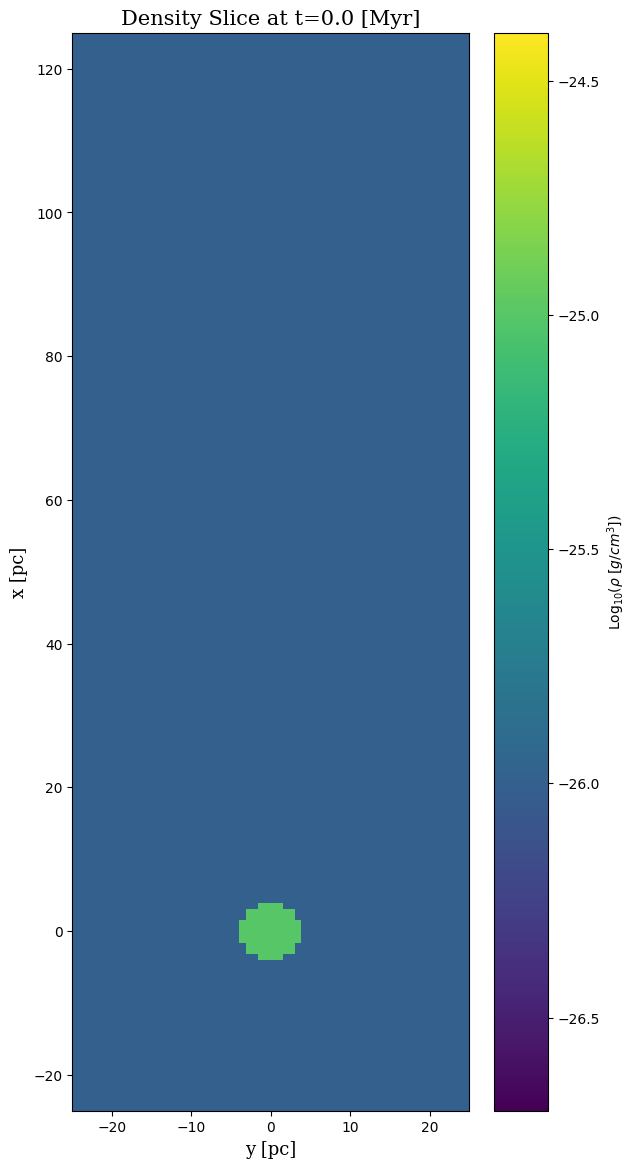}
        \caption{Sharp edges cloud.}
        \label{fig:sharp_cloud}
        \end{subfigure}
        \begin{subfigure}[b]{0.3\linewidth}
        \includegraphics[width=\linewidth]{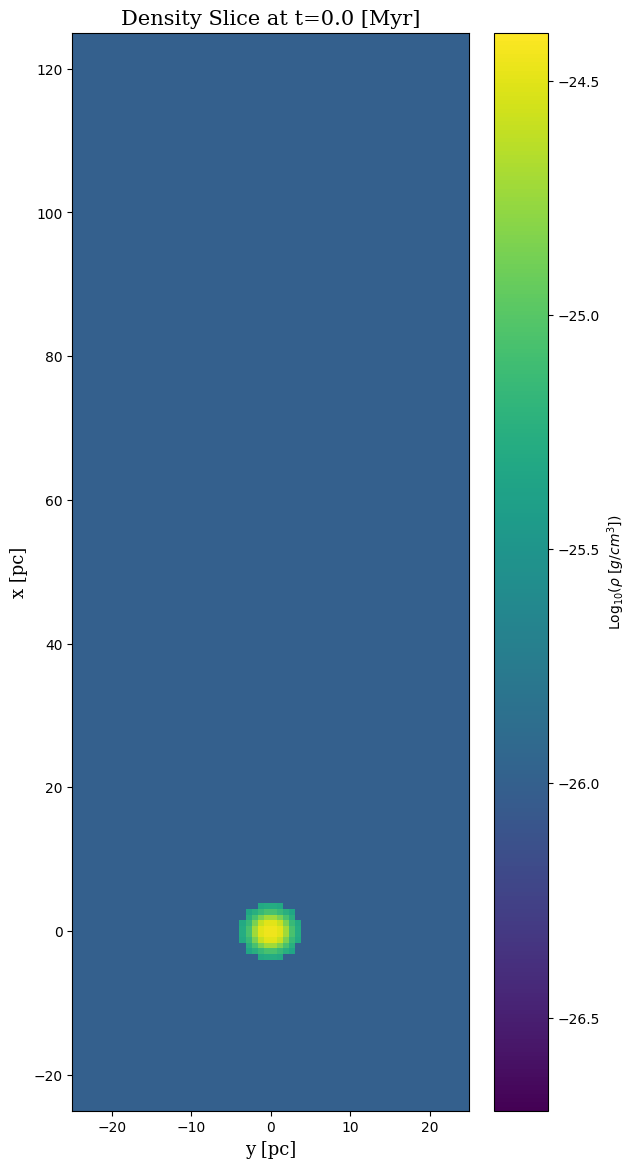}
        \caption{Smooth edges cloud.}
        \label{fig:smooth_cloud}
    \end{subfigure}
    \caption{2D Density slices on the XY plane at $z=0$ and $\rm t=0.0$ Myr of the two wind-cloud models.\it}
    \label{fig:ic-WC}
\end{figure}

For wind-cloud collisions, we simulate 2 models. Both are spherical with a radius of 4 parsec. The first model includes a completely symmetrical sphere with sharp edges (i.e. with a uniform density distribution) $\rho (r) = \rho _c$ for $r<=r_c$; on the other hand, the second model includes a sphere but with a smoothed density distribution (i.e. dense in the core and gradually dispersed towards the edges), modeled by: 
\begin{equation}
    \rho(r)=\rho_{\mathrm w} + \frac{(\rho_{c} - \rho_{\mathrm w})} {1 + \left( \frac{r}{r_{\mathrm{core}}}\right)^N}
    \label{eq:smooth}
\end{equation}
where $\rho_c$ is the density at the centre of the cloud, $\rho_w$ is the density of the wind, $r_{\rm core}$ is the radius of the cloud core, and N is an integer that determines the steepness of the curve describing the density gradient \cite{Nakamura et al.(2006)}. The wind that interacts with the cloud is on the X-axis with a velocity of 5 $\,\rm km\,s^{-1}$. The initial conditions are summarized in Table \ref{tab:ic-WC} and graphically in Figure \ref{fig:ic-WC}.

\begin{table} [h!]
    \centering
    \begin{tabular}{l c c c c c c} \br
        Model      & Resolution & Number of cells           & $\gamma$ & $\chi$ & r$_{cloud}$ & $v_{wind}$ \\ 
                   &            &                           &          &        & (pc)        & ($\,\rm km\,s^{-1}$)     \\  \hline    
        sph-sharp  & R$_5$      & $192 \times 64 \times 64$ &  $5/3$   & $10^2$ & 4.0         & 5.0        \\
        sph-smooth & R$_5$      & $192 \times 64 \times 64$ &  $5/3$   & $10^2$ & 4.0         & 5.0        \\ \br
    \end{tabular}
    \caption{Initial conditions for wind-cloud collisions.}
    \label{tab:ic-WC}
\end{table}

Our python-based shock-finding algorithm, designed to detect hydrodynamical shocks in numerical simulations, is the velocity jump (VJ) method, which is based on previous works of Vazza et al. (2011) \cite{2011MNRAS.418..960V}, Lehmann et al. (2016) \cite{2016MNRAS.463.1026L}, and Teutloff (2021) \cite{teutloff2021shock}. Our program searches for convergent flows ($ \Vec{\nabla}\cdot \Vec{v} < 0$) and pressure gradients larger than a  threshold $\Pi$ ($\lvert\Vec{\nabla p }\rvert > \Pi$), where the derivatives can be approximated by the central differences method. 


The simulation data files (in VTK format) contain several physical parameters such us, gas density ($\rho$), thermal pressure ($P$), velocity field $\vec{v}$ components ($v_x, v_y, v_z$) and cloud tracers (tr). The data of the different variables are then saved as 3D arrays to calculate other variables from the input variables, such as temperature  $T = \frac{P}{\rho}$, sound speed, and speed $|\Vec{v}|$. All variables are in code units. Then, we need to multiply by the normalization units. After normalizing the different variables, our python routine computes integrated quantities and searches for shocked cell candidates. The general steps that our shock finder code uses are:

\begin{itemize}
    \item Read the VTK files.
    \item Calculate the velocity divergency $\vec{\nabla} \cdot \vec{v} < 0$
    \item Calculate the pressure gradient $\lvert\Vec{\nabla p }\rvert > \Pi$
    \item Tag shock wave candidate cells that satisfy both conditions using cloud trackers.
    \item Calculate the sound speed in each cell $c_{s}=\sqrt{\gamma\frac{P}{\rho}}$
    \item Find the minimum $c_{s}$ in nearby cells.
    \item Calculate the Mach number in each cell.
    \item Combine Mach numbers and tagged cells.
\end{itemize}


\section{Results}
\label{results}
\subsection{Cloud-Cloud Collisions} 
 The evolution of CCc simulations can be split into four stages: a) Pre-collision stage, b) Compression stage, c) Pass-through stage, and d) dissipation stage. The pre-collision stage, similar to both models, is characterized by the motion of the moving cloud in a collision trajectory and the interaction of the cloud boundaries with their surrounding medium, creating Kelvin-Helmholtz instabilities. The motion also produces a relatively low-density and pressure region known as the rarefaction zone.
 
 The first interaction of both clouds occurs in the compression stage. At this moment, the moving cloud compresses the more external cloud, creating internal shocks on the initially non-moving cloud and its front. Also, there is an increase in temperature and pressure in the non-moving cloud product of this interaction. A characteristic U shape is shown in the larger cloud. In the next stage, the moving cloud passes through the initially stationary cloud, forming a bow shock and back-flows in the rarefaction zone from the stripped material of both clouds. This interaction promotes extreme values in variables such as pressure and density, with the strongest shocks observed along the bow shocks. The U-shaped structure in the stationary cloud is now extended and deformed, with a pronounced Mach cone visible at the rear of the moving cloud. Finally, in the dissipation stage, the collided gas of both clouds propagates and interacts with the ISM, and behind the bow shock. The Mach cone is evident in low-density regions. Turbulent dissipation at small subsonic eddies becomes the dominant process over shock dissipation.

 \begin{figure}[!htbp]
  \centering
  \includegraphics[scale = 0.09]{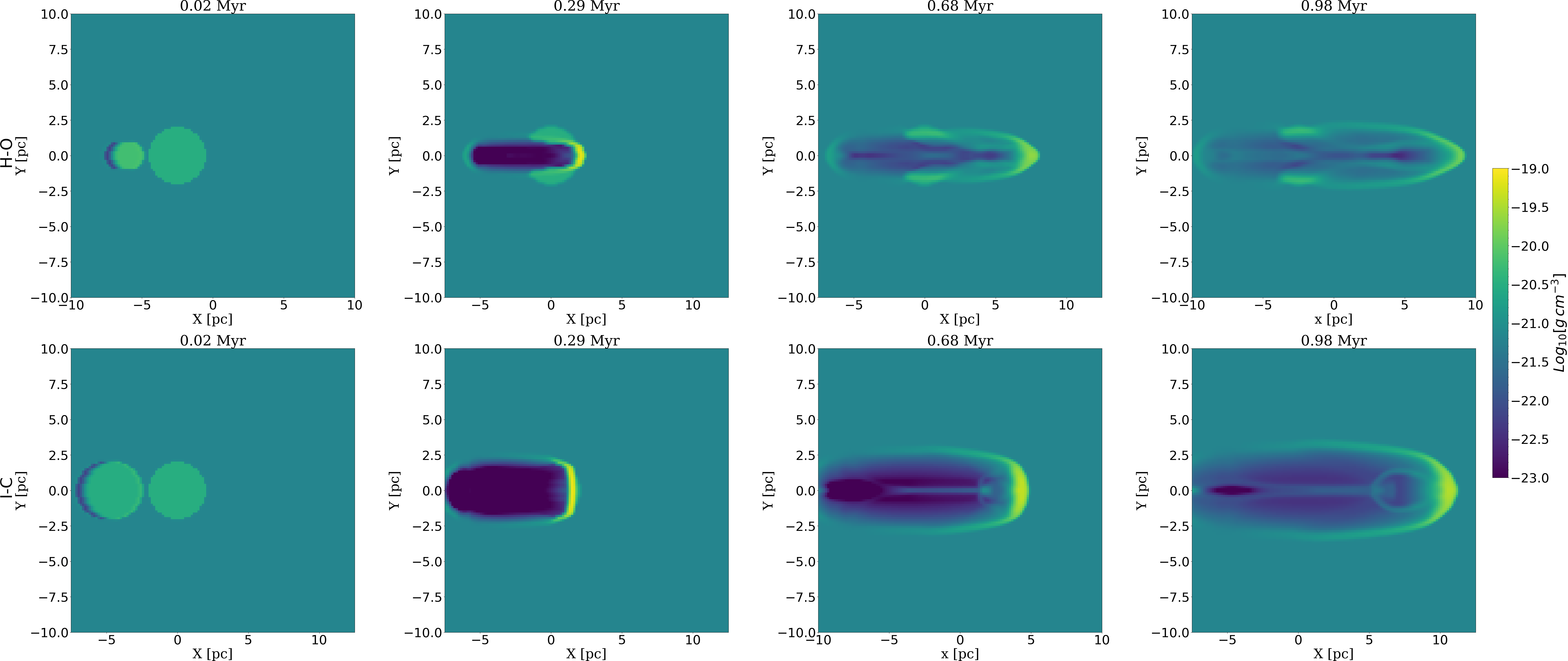}
  \caption{Density maps on different times. The color bar is on a logarithmic scale. The top row shows the H-O model, while the bottom row shows I-Cmodel}
  \label{fig:flow_chart}
\end{figure}
 The main difference between the two models is the cross-section, which is larger for the I-C model, resulting in a larger amount of gas affected by the collision. Also, this model presents a more pronounced U shape than the H-O model, with in turn has a narrow U shape and a thin foot point. Both models have the presence of wakes. H-O shows thick and dense wakes, while the I-C model has more diffusive wakes as a product of  dissipation. The Mach cone is more evident in the I-C model and has a lower density in the rarefaction zone. Both models destroy the clouds. 

\begin{figure}[!htbp]
  \centering
  \includegraphics[scale = 0.5]{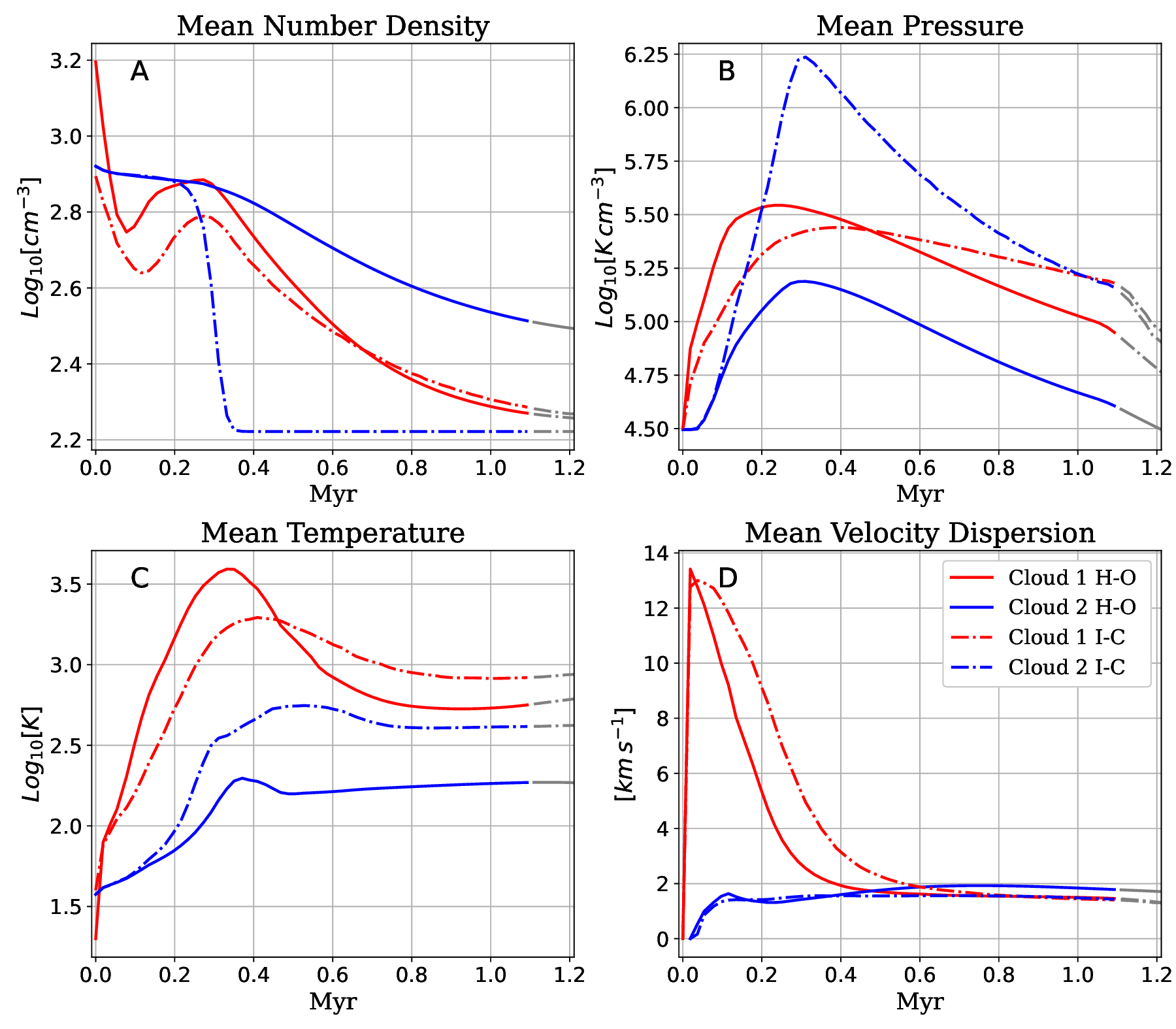}
  \caption{Diagnostic curves of the moving cloud in red and the initially non-moving cloud in blue in our two CCc models for varying sizes of the moving cloud. The number density $n$ is shown in panel A, the pressure in panel B, the temperature in panel C, and the velocity dispersion in panel D. Labels are presented in panel B, where the H-O model is shown with solid lines, and the I-C model is shown with dashed lines.}
  \label{fig:Diag_cc}
\end{figure}

 The diagnostic curves of our CCc models are presented in Figure \ref{fig:Diag_cc}, which shows with dashed lines the identical cloud model and with solid lines the H-O model (in both cases, the collision velocity is $30 \,\rm km\,s^{-1}$), revealing differences in the evolution of key physical properties over time. In the I-C model, the number density presents a significant drop for the initially non-moving cloud, suggesting its dissipation while the moving cloud remains dense. The I-C  model also exhibits a greater increase in pressure and temperature in the initially non-moving cloud. In contrast, the moving cloud in the H-O model shows higher values due to interaction with the gas in the U cavity.  Velocity dispersion indicates that turbulence produced by the collision is more evident for the I-C model with higher $\rm{M_{rsm}}$, but when it reaches subsonic values, turbulence does not return to supersonic as the H-O model.

\begin{figure}[!htbp]
  \centering
  \includegraphics[scale =0.42]{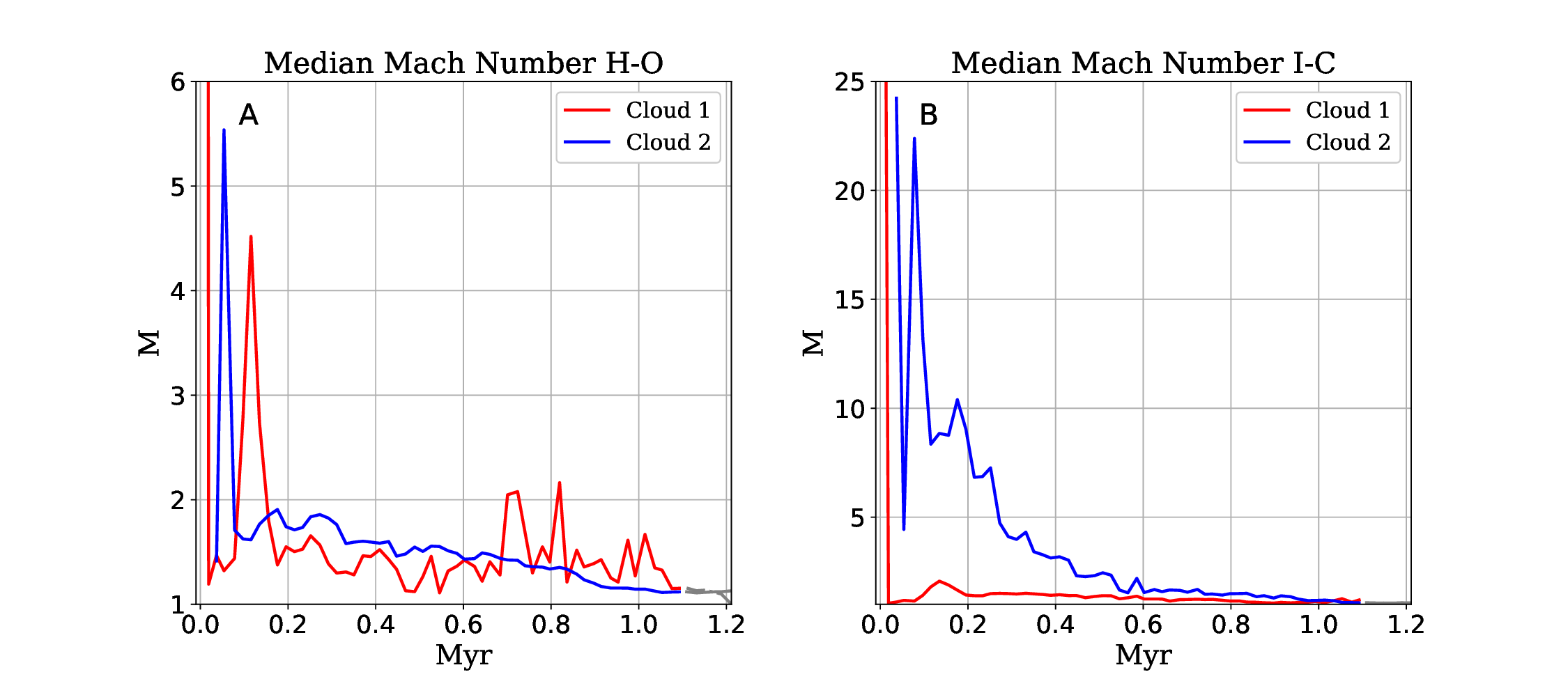}
  \caption{Time evolution of the median Mach number in panel A (H-O model), and in panel B (I-C model).}
  \label{fig:Mach_cc}
\end{figure}

 The behavior of both models is almost the same regarding shock distribution. Stronger shocks can be found in the front layers of the collision, along the bow shock, and alsoat the back of the rarefaction region. Another important zone where the shocks can be found is in the Mach cone, produced by the reflected shocks. Figure \ref{fig:Mach_cc} shows the time evolution of the median Mach number (the H-O model is in panel A and the I-C model is in panel B). Both models show that strong shocks occur at the early stages of the collisions. Moreover, the main difference is that the shocks in the I-C model are stronger than in the H-O model. Despite these differences, both models converge to similar median Mach numbers at the end of the simulations. Another important remark is that there are more shocks in the I-C model in the initially non-moving cloud as a result of the large amount of gas affected.
 
\begin{figure}[!htbp]
  \centering
  \includegraphics[scale =0.5]{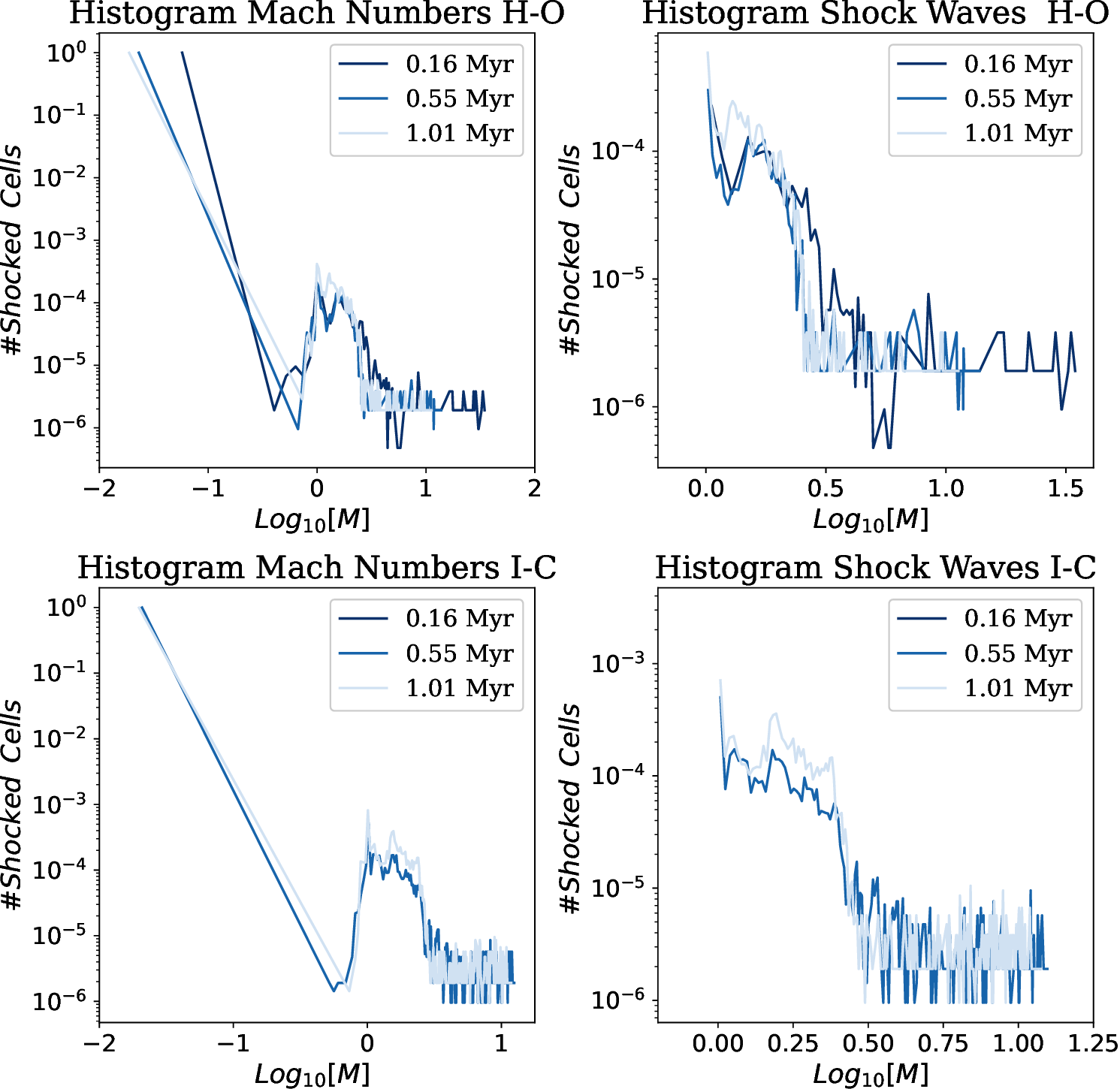}
  \caption{Histograms showing the Mach numbers for the moving cloud of all of the subsonic and supersonic perturbations (left) and only of supersonic shocks (right). }
  \label{fig:both_models}
\end{figure}

Figure \ref{fig:both_models} illustrates the evolution of the shock population over time in both models. At the beginning of the simulations, higher Mach numbers are observed. As time evolves, the shock Mach number and the number of shocked cells decrease until they reach and stabilize around a certain Mach number $\sim 3$ for both models. Notice that the left panel shows the histograms with the total Mach number considering subsonic and shock waves. Most perturbations are subsonic waves; however, the amount of shock waves is also significant, especially for the I-C model. As time evolves, there is a shift toward small Mach numbers, suggesting that energy dissipation is predominantly controlled by turbulent eddies rather than shocks, corroborating findings from turbulence studies by e.g. \cite{2009ApJ...692..364F}.

These models have some important physical implications. First, for the star formation processes that can be triggered by CCc. Our Mach number analysis indicates that star formation can be initiated at the early stages of the collision, given that stronger shocks are produced at those times. Strong shock can modify the physical and chemical composition of the gas through heating and compression. Shock heating is important regardless of the CCc model considered (see panel C in figure 4), but it is $\sim 1 dex$ more significant in the H-O model. Another implication is the influence of the cloud radius.  Although the I-C model is more idealised, it provides good insights into the physics of the collision. A direct relation exists between the strength of the shock and the size of the clouds, showing that clouds of relatively the same size are more likely to trigger star formation, owing to the extra compression to which gas is subjected (see panel B in figure 4).

Another physical implication of CCc is the chemistry changes that heating can induce in the gas resulting from the collision. Shocks increase the temperature as a product of converting the kinetic energy of the gas into thermal energy. It also causes the molecular gas inside the cloud to become warmer, which can modify the chemistry of the molecules in the cloud \cite{2008ApJ...689..865K}.

\subsection{Wind-Cloud Collisions} 

\begin{figure}[!htbp]
    \centering
    \includegraphics[width=0.75\textwidth]{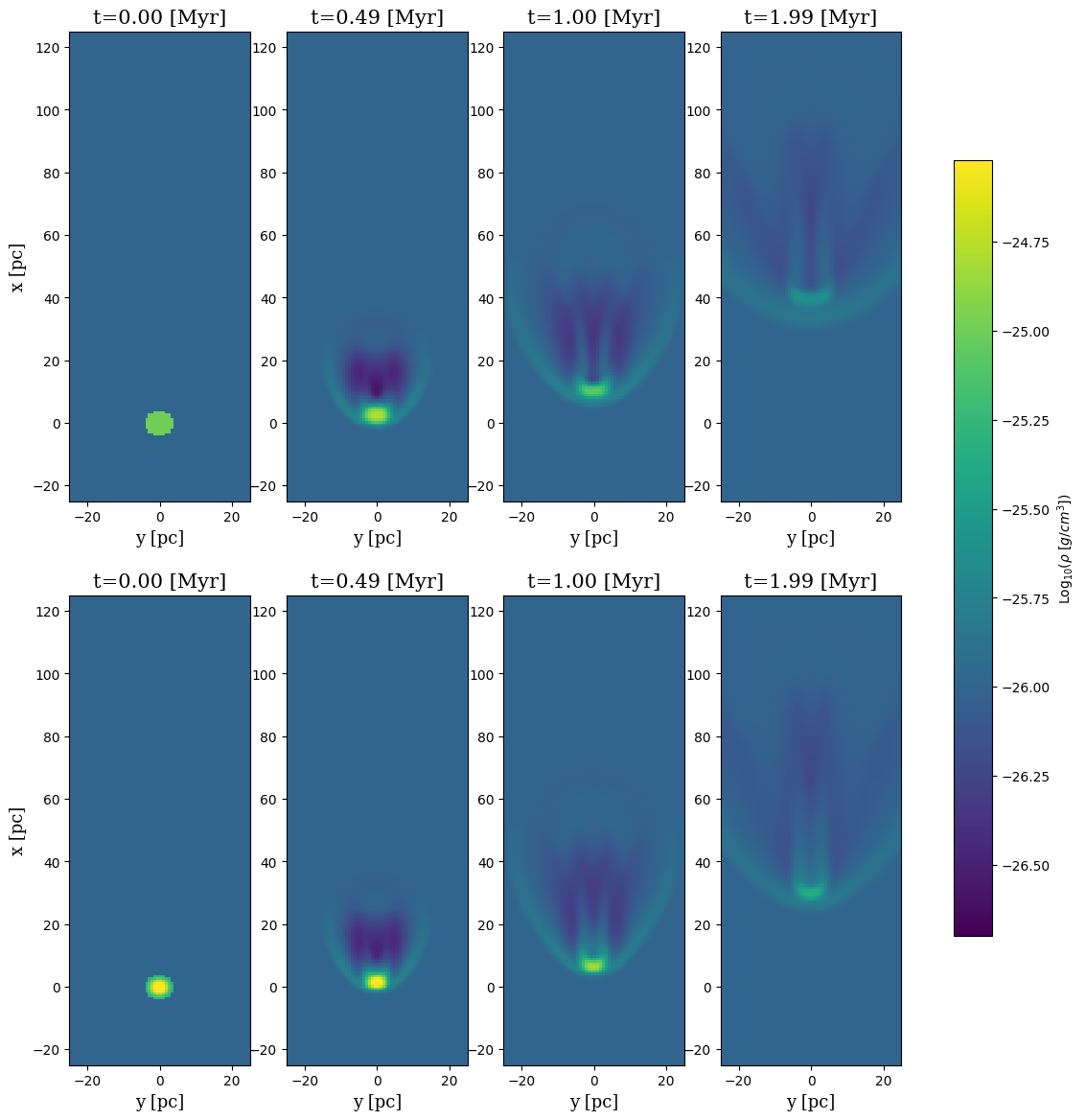}
    \caption{2D slices at $z$=0, showing the evolution of the logarithm of the density in the smoothed-edges cloud (top) and the sharp-edges cloud (bottom) models. }
    \label{fig:wc_compile}
\end{figure}

In this type of interaction, the cloud is destroyed over time by the wind (Figure \ref{fig:wc_compile}), due to the difference in density, pressures and instabilities that arise from the relative motion between the cloud and the wind. Then, similar to the work by \cite{2016MNRAS.455.1309B}, we also found that the evolution of WCc occurs in 4 phases: compression phase, stripping phase, expansion phase, and break-up phase. In the initial phase, shocks form in both the wind and the cloud. The cloud experiences inward shocks and an external bow shock emerges. The pressure gradient forces of the wind compress the cloud, increasing the core's density. During the stripping phase, the cloud material is transported downstream by the wind, resulting in the formation of an extended tail aligned with the wind direction. This phenomenon is induced by Kelvin-Helmholtz (KH) instabilities. The gas then enters an expansion phase, due to the passage of internal shocks that expand the cloud, making it susceptible to dispersion. As the density of the cloud continues to decrease, it transitions into the break-up phase. In this last stages, the cloud undergoes fragmentation into smaller cloudlets as a result of Rayleigh-Taylor (RT) instabilities, and these smaller entities are accelerated more rapidly downstream.

The KH and RT instabilities can also explain the increase in kinetic energy and temperature in the last phase of the shock. As the cloud is breaking, turbulence is generated inside the cloud. Then, filaments in the borders of the cloud emerge, accelerating the process of disruption, and as a consequence the kinetic energy rises. Furthermore, the resulting changes in thermal pressure lead to heating, and the cloud expands more rapidly.

\begin{figure}[!htbp]
    \centering
    \includegraphics[width=0.80\textwidth]{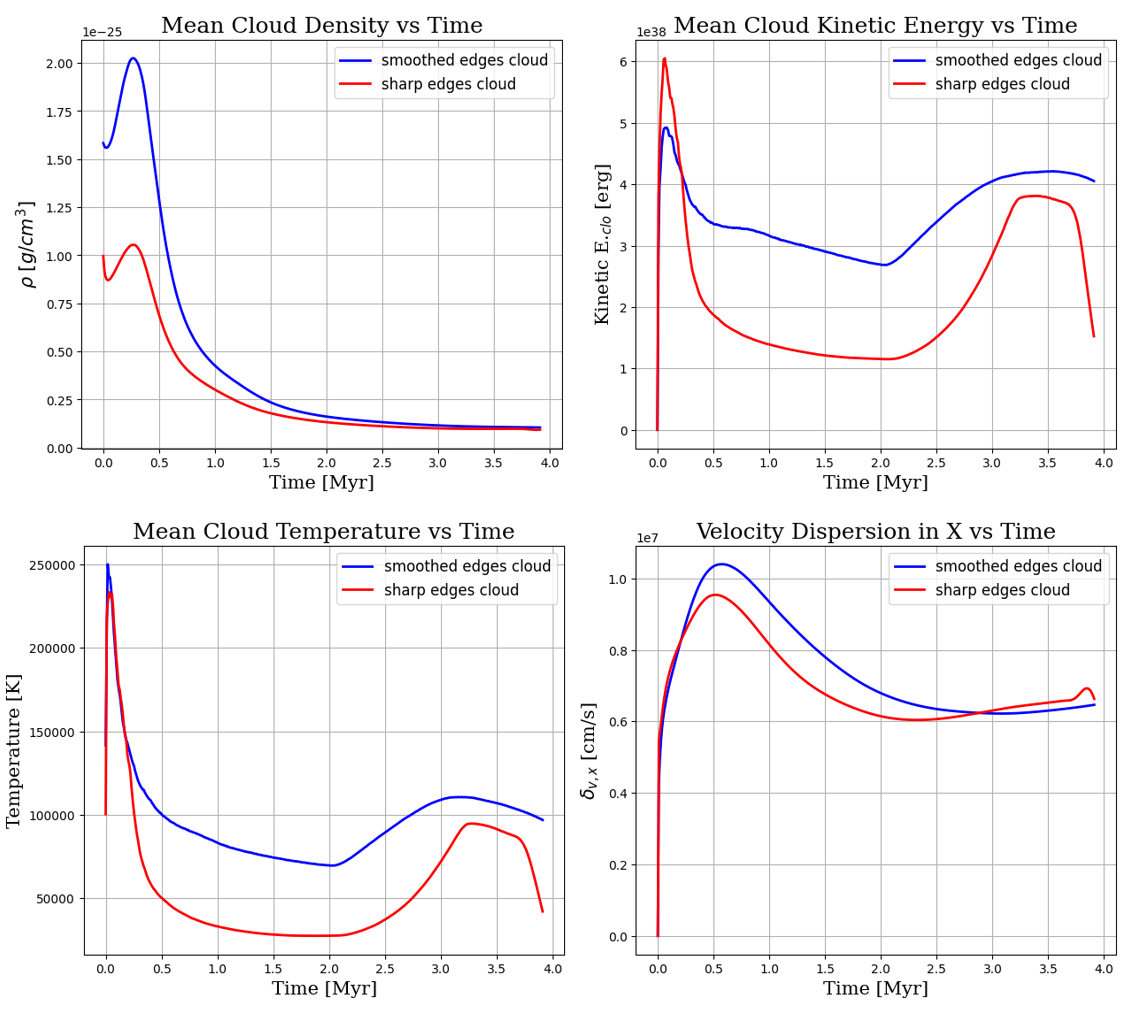}
    \caption{Diagnostic curves of the WCc models in time. The mean cloud density is shown in panel A, the mean kinetic energy in panel B, the temperature in panel C, and the velocity dispersion in panel D.}
    \label{fig:wc_diagnostic}
\end{figure}

Something to note is that in the top-left panel of Figure \ref{fig:wc_diagnostic}, we can appreciate that the cloud with smoothed edges has a considerably higher initial density than the other model. This happens because a correction factor was added to the density so that both clouds have the same initial mass.
Although the dispersion speed is similar in both models, it is evident that in the cloud with smoothed edges, it is slightly higher during the stripping stage. This is because the layer that covers the cloud disperses quickly due to K-H instabilities and turbulence. 
Something similar happens with the kinetic energy. Since the outermost layer of the cloud with smoothed edges dissipates faster, it reaches a higher speed, so the kinetic energy also increases.
Regarding temperature, in both models the temperature increases abruptly at the beginning due to the intense collision between the cloud and the wind, which causes the material to compress first, then expand, and also heat up. The temperature is higher in the smoothed cloud see Figure \ref{fig:wc_diagnostic}. This is because clouds with smoothed edges are more prone to dynamical instabilities, which influence mixing. If cloud gas is more mixed with the wind, its temperature becomes higher.

\begin{figure}[!htbp]
    \centering
    \includegraphics[width=0.50\textwidth]{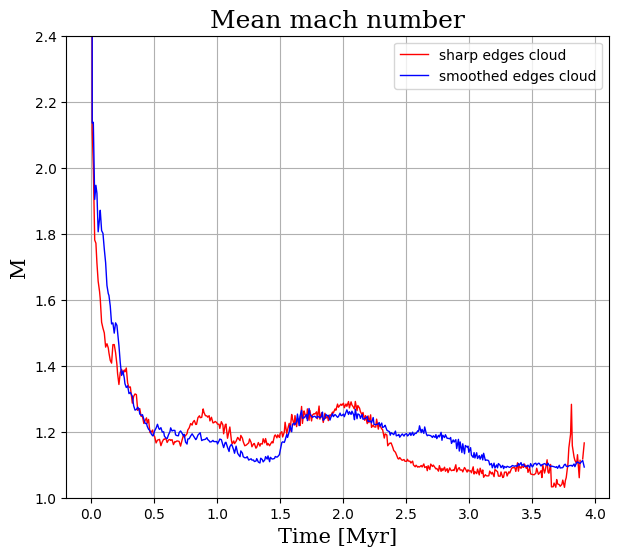}
    \caption{Mean cloud mach numbers for wind-cloud collisions.}
    \label{fig:wc_mean_mach}
\end{figure}

As Figure \ref{fig:wc_mean_mach} shows, the distribution of shocked cells in the cloud remains relatively consistent over time, remaining below of M $<$ 2.5. for both models. Thus, there is little influence of cloud geometry on the resulting distribution of shocks. However, during the compression phase, stronger shocks occur, triggered by the initial collision between the moving wind and the stationary cloud. The population of strong shocks decreases throughout the stripping and expansion stages as a result of the lateral dispersion of the gas and the weakening of compression (which primarily occurs during the initial stages).

\begin{figure}[h!]
    \centering
    \begin{subfigure}[b]{0.45\linewidth}
        \includegraphics[width=\linewidth]{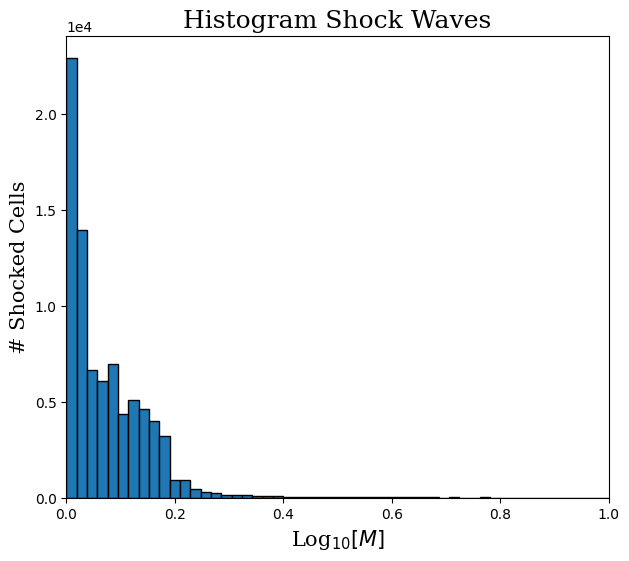}
        \caption{Sharp edges cloud.}
        \label{fig:/wc_histo_schok_sharp}
        \end{subfigure}
        \begin{subfigure}[b]{0.45\linewidth}
        \includegraphics[width=\linewidth]{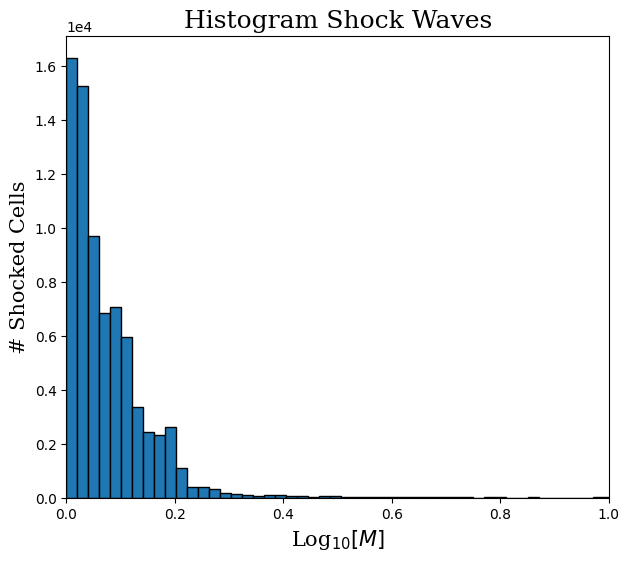}
        \caption{Smooth edges cloud.}
        \label{fig:wc_histo_schok_smoth}
    \end{subfigure}
    \caption{Histograms showing the Mach number of supersonic shocks in wind-cloud collisions.}
    \label{fig:wc_histo_schok}
\end{figure}

In the shock histograms (see Figure \ref{fig:wc_histo_schok}), a similar log-normal distribution is observed in both models. However, there are some small differences. We observe that there are a larger number of shock cells near $M\sim 1$ in the sharp edges cloud (Figure \ref{fig:/wc_histo_schok_sharp}), and the tail of the log-normal distributions is less pronounced.

Phases exert notable influences on diagnostics, including cloud density, kinetic energy, temperature, and dispersion. Shock waves are important for heating up cold clouds in the interstellar medium, as they are embedded in hot stellar-driven outflows. Shocks inject thermal energy into the gas through which they pass. Compared to strong shocks, low Mach number shocks are more numerous, so they heat up larger amounts of gas.

The compression and heating of cold clouds in the ISM by shock waves has implications for star formation, chemical evolution, and ISM dynamics. 
For example, the heating can increase the pressure and turbulence of the clouds, which can influence their gravitational stability and their ability to collapse and form stars \cite{2014MNRAS.444.2884L}. It can also alter the chemical composition of the clouds, by triggering reactions between molecules and atoms, or by dissociating molecules by the ultraviolet radiation generated by the shock waves. Moreover, wind-cloud interactions can modify the structure and motion of the ISM, by creating bubbles, filaments, and multi-cloud interactions that influence the distribution and transport of matter and energy out of galaxies \cite{Villares_etal2024}.

\newpage
\section{Conclusions}
\label{conclusions}

We have studied two types of collisions in the ISM, cloud-cloud (CCc) and wind-cloud (WCc). Both are quite universal and appear in various scenarios, which are relevant for understanding the ISM dynamics, chemistry, SF history, and turbulence. We determine that:

\begin{itemize}
    \item The dynamics evolution of the CCc can be separated into 4 stages: \textbf{ a) Pre-collision stage}. The moving cloud is in a collision trajectory,\textbf{  b) Compression stage}. Gas compression of the initially non-moving cloud,  \textbf{c) Pass-through stage}. The moving cloud had passed through the non-moving cloud creating internal shocks and a Mach cone, and\textbf{   d) Dissipation stage}. Gas of both clouds interacts with the medium and the energy dissipation changes from shocks to turbulent eddies.
    \item The impact of varying the size of the cloud directly affects the amount of gas affected by the collision due to the increase in cross-section and in the creation of shocks with higher Mach numbers. The integrated quantities also have an impact because, for the H-O model, the moving cloud is more affected, given that it is enveloped by the gas of the non-moving cloud as it passes through it.
    \item Regarding shock Mach numbers, there is a scaling effect with time. The shock distribution evolves from stronger Mach numbers at early stages to weaker shocks with smaller Mach numbers at late stages. Most shocks populate the bow shock and the rarefaction region. Even though the simulation generates mostly subsonic waves, there is a significant amount of strong shocks. 
    \item The implication for the physics of the ISM is that the CCc modify the gas structure of clouds, changing their physical and chemical properties such as temperature, pressure, and densities. Gas compression in CCc can also trigger star formation at the early stages of the collision because strong shocks are produced at those times.
    \item We also found that the shape of the clouds in WCc affects the dynamics of the evolution of integrated quantities in different stages. The smoothed-edge cloud has higher kinetic energy and temperature than the sharp-edge cloud. This is because the smoothed cloud is more prone to RT and KH instabilities, which generate turbulence and mixing.

\end{itemize}

Our results show that strong shocks are ubiquitous in ISM collisions and play an important role in regulating gas heating, and cloud structure, which play a role in regulating star formation. Future work includes implementing a GPU-based code, studying the effects in different shapes of clouds, and adding other physical properties such as self-gravity, turbulence, and magnetic fields.

\section*{Acknowledgments}
Part of the material presented here is based on the thesis work by S. Navarrete at Universidad Yachay Tech (\cite{4480}). The authors gratefully acknowledge the Gauss Centre for Supercomputing e.V. (\url{www.gauss-centre.eu}) for funding this project by providing computing time (via grant pn34qu) on the GCS Supercomputer SuperMUC-NG at the Leibniz Supercomputing Centre (\url{www.lrz.de}). In addition, the authors thank CEDIA (\url{www.cedia.edu.ec}) for providing access to their HPC cluster as well as for their technical support. The authors also thank the developers of the PLUTO code for making this hydrodynamic code available to the community.


\end{document}